\documentclass[reprint,aps,prl,showpacs,showkeys]{revtex4-1}

\usepackage{natbib}
\usepackage{graphicx}
\usepackage{bm}
\usepackage{hyperref}

\begin{document}

\title{Exact Axisymmetric Solutions of the Maxwell Equations\\
in a Nonlinear Nondispersive Medium}
\author{E.\,Yu.\,Petrov}
\author{A.\,V.\,Kudrin}
\email{kud@rf.unn.ru}
\affiliation{University of Nizhny Novgorod,
23 Gagarin Ave., Nizhny Novgorod 603950, Russia}
\begin{abstract}
The features of propagation of intense waves
are of great interest for theory and experiment
in electrodynamics and acoustics. The behavior of
nonlinear waves in a bounded volume is of especial importance
and, at the same time, is an extremely complicated problem.
It seems almost impossible to find a rigorous solution to
such a problem even for any model of nonlinearity. We obtain the
first exact solution of this type. We present a new method for
deriving exact solutions of the Maxwell equations in a nonlinear medium
without dispersion and give examples of the obtained solutions
that describe propagation of cylindrical electromagnetic
waves in a nonlinear nondispersive medium and free
electromagnetic oscillations in a cylindrical cavity resonator filled
with such a medium.
\end{abstract}


\maketitle

Wave propagation in nonlinear media is a fundamental and wide-ranging
problem in physics~\cite{Whi,Abl}. The possible self-steepening and
formation of shock discontinuities in large-amplitude pressure waves
is well known in fluid mechanics, being a typical nonlinear phenomenon.
A similar phenomenon (formation of surfaces of discontinuity for the
electric and magnetic fields) can also be observed during propagation
of intense electromagnetic waves in certain media, and there is an
elegant physical analogy between the fluid mechanics and electrodynamics
in this case. The discovery of materials with well-pronounced nonlinearity
of electromagnetic properties (for example, ferrites and ferroelectrics) has
attracted considerable attention to essentially nonlinear electromagnetic
phenomena, and some important advances have been made~\cite{Gap,Ros}. Further
theoretical progress, however, has met serious difficulties because such
phenomena cannot be described satisfactorily by perturbation theory,
and rigorous solutions of field equations are required in order to get
theoretical predictions. In view of the above, finding new,
physically important exact solutions of nonlinear partial differential
equations (PDEs) that describe the behavior of waves in nonlinear
media is very topical~\cite{Pol,Fok}. In most papers on the subject, plane
nonlinear waves are considered. At the same time, features of propagation
of nonlinear cylindrical and spherical waves, as well as the properties
of the related nonlinear PDEs in the corresponding curvilinear coordinates,
remain poorly studied.

In what follows, we present a new method for constructing exact
axisymmetric solutions
of the Maxwell equations in a nonlinear nondispersive medium.
It is assumed that the medium considered lacks a center of
inversion and the dependence of the electric displacement on the electric
field can be approximated by an exponential function.

Consider electromagnetic fields in a loss-free nonmagnetic medium.
With a view to analyzing uniaxial crystals, we assume that
the medium possesses an axis of symmetry, hereafter taken as the $z$ axis
of a cylindrical coordinate system ($r$, $\phi$, $z$). If the fields
are independent of $\phi$ and $z$, the Maxwell equations admit solutions
in which only the $E_z$ and $H_\phi$ components are nonzero ($E$ waves with
respect to the symmetry axis). Restricting ourselves to
consideration only of such solutions, we will also neglect dispersion
effects and suppose that the relation of the displacement $D_z$ to the
electric field $E_z$ is local in space and time. Denoting
$E_{z}(r,t)$, $D_{z}(r,t)$, and $H_{\phi}(r,t)$ as
$E$, $D$, and $H$, respectively, we can write equations for these
functions in the form
\begin{equation}
\partial_{r}H+r^{-1}H=\varepsilon(E)\,\partial_{t}E,\quad
\partial_{r}E=\mu_{0}\,\partial_{t}H,\label{eq1}
\end{equation}
where $\varepsilon(E)=dD/dE$. System~(\ref{eq1}) can be reduced to the
nonlinear wave equation
\begin{equation}
\partial^{2}_{r}E+r^{-1}\partial_{r}E=\mu_{0}\,\partial_{t}
(\varepsilon(E)\,\partial_{t}E).\label{eq2}
\end{equation}
It will be shown below that system~(\ref{eq1})
and Eq.\,(\ref{eq2}) are integrated exactly if
the function $\varepsilon(E)$ is chosen in the form
\begin{equation}
\varepsilon(E)=\epsilon_{0}\varepsilon_{1}\exp(\alpha E),\label{eq3}
\end{equation}
where $\varepsilon_1$ and $\alpha$ are certain constants. The
longitudinal component of the electric displacement can
be represented as $D=D_{0}+\alpha^{-1}\epsilon_{0}\varepsilon_{1}\left[\exp(\alpha E)-1\right]=D_{0}+\epsilon_{0}\varepsilon_{1}(E+
\alpha E^{2}/2+\ldots)$,
where $D_{0}=D(E=0)={\rm const}$.
It is clear that function~(\ref{eq3}), as any theoretical model of nonlinearity, cannot be used in the entire range $0<|E|<\infty$. However, for moderately small electric fields observed in actual experiments  ($|\alpha E|\ll 1$), the chosen dependence $D(E)$ correctly describes
dielectric properties of certain media with accuracy up to terms of order $E^2$ inclusively.
Since even powers of $E$
are present in the series expansion of $D$, the medium for which the
dependence $\varepsilon(E)$ is approximated by function~(\ref{eq3}) does not possess
a center of inversion~\cite{Fra}. This is inherent in, e.g., uniaxial pyroelectric
and ferroelectric crystals, provided that the $z$ axis is aligned with
the crystallographic symmetry axis. The case where $D_{0}\neq 0$ corresponds
to the presence of spontaneous polarization. The value of $\alpha$ can be obtained, for example, in experiments on microwave frequency doubling in ferroelectric crystals~\cite{Iva,Bel}. Along with
the medium properties, another factor leading to lack of a center of inversion can be
the presence of a strong external electric field~\cite{Fra}. For example, let an
isotropic medium in which $D=\epsilon_{0}\varepsilon_{1}E_{z}+\chi^{(3)}E^{3}_{z}$,
where $\chi^{(3)}={\rm const}$~\cite{Ros}, be placed in a uniform static electric field
${\bf E}=E_{0}{\hat {\bf z}}_{0}$.
Representing the total field as $E_{z}=E_{0}+E$, we obtain
the term proportional to $E^{2}$ in the expansion of $D$. Thus, with
appropriately chosen constants $D_{0}$, $\varepsilon_{1}$, and $\alpha$,
formula~(\ref{eq3}) correctly describes dielectric properties of media lacking a
center of inversion in the case of weak nonlinearity where we can
restrict ourselves to the quadratic (in $E$)
correction term
to the linear dependence of $D$ on $E$.

Note that the forthcoming
results can also be related to magnetic media lacking a center of
inversion, such as ferromagnetic crystals. To this end, one should consider
$H$ waves, for which $H_{z}\neq 0$ and $E_{\phi}\neq 0$, and put
$dB_{z}/dH_{z}=\mu_{0}\mu_{1}\exp(\nu H_{z})$.

Let us use the following ansatz in system\,(\ref{eq1}):
\begin{equation}
E=\alpha^{-1}(u-2\xi),\quad
H=\varepsilon^{1/2}_{1}(Z_{0}\alpha)^{-1}\,e^{-\xi}(v-2\eta),\label{eq4}
\end{equation}
where $\xi={\rm ln}(r/r_{0})$,
$\eta=t(\epsilon_{0}\varepsilon_{1}\mu_{0})^{-1/2}/r_{0}$,
$Z_{0}=(\mu_{0}/\epsilon_{0})^{1/2}$, and $r_0$ is an arbitrary constant
with the dimension of length. In the new variables, we have
\begin{equation}
\partial_{\xi}u=\partial_{\eta}v,\quad
\partial_{\xi}v=e^{u}\,\partial_{\eta}u.\label{eq5}
\end{equation}
System~(\ref{eq5}) has particular solutions in which one of the functions
$u$ and $v$ can be expressed in terms of the other:
\begin{equation}
u=F(\xi \pm \eta\,e^{-u/2}),\quad
v=\pm 2(e^{u/2}-1).\label{eq6}
\end{equation}
Here, $F$ is an arbitrary differentiable function. Similar solutions,
which are analogous to the Riemann solutions in fluid
mechanics, have been obtained in~\cite{Gap}. However, it can easily be verified that ansatz~(\ref{eq4}) does not
make it possible to arrive at physically admissible solutions for $E_z$
and $H_\phi$ on the basis of~(\ref{eq6}) in our case~\cite{Kud},
and somewhat another approach should be used.
The approach is based on the application of a
hodograph transformation for seeking solutions for which the Jacobian
$D(u,v)/D(\xi,\eta)$ is nonzero. Using $u$ and $v$ as
independent variables, we obtain from~(\ref{eq5}) the system of linear
equations
\begin{equation}
\partial_{v}\eta=\partial_{u}\xi,\quad
\partial_{u}\eta=e^{u}\,\partial_{v}\xi.\label{eq7}
\end{equation}
Excluding $\eta$ from~(\ref{eq7}) yields the equation
$\partial^{2}_{u}\xi=e^{u}\partial^{2}_{v}\xi$, which, by making the replacement $w=2e^{u/2}$, reduces to
\begin{equation}
\partial^{2}_{w}\xi+w^{-1}\partial_{w}\xi=\partial^{2}_{v}\xi.\label{eq8}
\end{equation}
A remarkable symmetry property of system~(\ref{eq1}) with exponential
nonlinearity is that it is reduced
to a linear wave equation of form~(\ref{eq8}) for cylindrical waves
by application of the above-described substitutions
and the hodograph transformation. However, initial and boundary
conditions for the fields $E$ and $H$ in the new variables $\xi$
and $\eta$ can become much more complicated. This may cause
the necessity of numerically solving even linear equation~(\ref{eq8}).
Nevertheless, it is possible to propose a comparatively simple analytical method
which permits one to find physically admissible exact
solutions of system~(\ref{eq1}). The idea of the method consists in the
following. At first, one should find an analytical solution
to the problem of propagation of cylindrical $E$ waves in a medium with
the linear dependence $D_{z}=D_{0}+\epsilon_{0}\varepsilon_{1}E_{z}$. Assume
that such a solution is known and we have the functions $E$ and $H$
satisfying the linear field equations and the specified initial and
boundary conditions. The characteristic spatial scale determined by
these conditions for the problem considered will be
denoted by $a$. We also introduce the dimensionless variables
$\rho=r/a$ and $\tau=t(\epsilon_{0}\varepsilon_{1}\mu_{0})^{-1/2}/a$.
Then it is convenient to represent the solution of
the linear problem in the form
\begin{equation}
E\equiv E_{z}={\cal E}(\rho,\tau),\quad
H\equiv H_{\phi}=Z_{0}^{-1}\varepsilon^{1/2}_{1}\,{\cal H}(\rho,\tau),\label{eq9}
\end{equation}
where the functions ${\cal E}$ and ${\cal H}$ satisfy the system
\begin{equation}
\partial_{\rho}{\cal H}+\rho^{-1}{\cal H}=\partial_{\tau}{\cal E},\quad
\partial_{\rho}{\cal E}=\partial_{\tau}{\cal H}.\label{eq10}
\end{equation}
We write the quantities $\xi$ and $\eta$ as
\begin{eqnarray}
&&\hspace{-5mm}
\xi=C_{1}{\cal E}(w,v)+{\rm ln}\,\frac{w}{2}\,,\;\;\;
\eta=\frac{C_{1}}{2}\,w\,{\cal H}(w,v)+\frac{v}{2}\,,\label{eq11}
\end{eqnarray}
where $C_1$ is an arbitrary constant.
It can easily be verified by straightforward differentiation
that functions~(\ref{eq11}) satisfy system~(\ref{eq7}). Using formulas~(\ref{eq4}),
we can pass to the initially used quantities $r$, $t$, $E$, and $H$ in~(\ref{eq11}).
Putting $C_{1}=-\alpha/2$ and $r_{0}=2a$ ensures
that the resulting solution will go into solution~(\ref{eq9})
in the linear case. Bearing this in mind,
after some simple algebra we obtain
\begin{eqnarray}
&&\hspace{-5mm}
E={\cal E}\left(\rho\,e^{\alpha E/2},
\tau+\alpha Z_{0}\rho H/(2\sqrt{\varepsilon_{1}})\right),\nonumber\\
&&\hspace{-5mm}
H=\frac{\varepsilon^{1/2}_{1}}{Z_{0}}\,e^{\alpha E/2}
{\cal H}\left(\rho\,e^{\alpha E/2},
\tau+\alpha Z_{0}\rho H/(2\sqrt{\varepsilon_{1}})\right).\label{eq12}
\end{eqnarray}
These expressions give an exact solution of system~(\ref{eq1})
in implicit form and describe axisymmetric electromagnetic fields
in the nonlinear medium considered. For the known functions
${\cal E}$ and ${\cal H}$, which are determined by solving the
linear problem, and given values of $\rho$ and $\tau$, formulas~(\ref{eq12})
represent a system of two transcendental equations in
$E$ and $H$. In the limit $\alpha\to 0$, the solution obtained goes
into solution~(\ref{eq9}) of the linear problem, but, generally, corresponds
to somewhat different initial or boundary conditions compared
with those satisfied by functions~(\ref{eq9}). Let us now discuss
some particular examples to better understand the essence of this method.

{\em Initial value problem.} Let the initial field distributions
\begin{equation}
\left.E_{z}\right|_{t=0}\equiv{\cal E}(\rho,0)=\beta(1+\rho^{2})^{-3/2},\quad
\left.H_{\phi}\right|_{t=0}\equiv 0,\label{eq13}
\end{equation}
where $\beta$ is a certain constant, be specified in
a linear medium with constant dielectric permittivity
$\varepsilon=\epsilon_{0}\varepsilon_{1}$. To find $E_z$ and $H_\phi$ for
$t>0$, we apply the Hankel transform and obtain a solution of system~(\ref{eq10}) under
initial conditions~(\ref{eq13}) as follows:
\begin{eqnarray}
{\cal E}(\rho,\tau)
&=&\beta\,{\rm Re}\left\{(1-i\tau)[(1-i\tau)^{2}+\rho^{2}]^{-3/2}
\right\},\nonumber\\
{\cal H}(\rho,\tau)
&=&\beta\rho\,{\rm Re}\left\{i[(1-i\tau)^{2}+\rho^{2}]^{-3/2}
\right\}.\label{eq14}
\end{eqnarray}
We now write the corresponding exact solution of nonlinear system~(\ref{eq1})
with $\varepsilon(E)$ in form~(\ref{eq3}). Substituting ${\cal E}$ and ${\cal H}$
given by~(\ref{eq14}) into~(\ref{eq12}), we have
\begin{eqnarray}
&&\hspace{-5mm}
E=\beta\,{\rm Re}\left\{(1-i\theta)[(1-i\theta)^{2}+\rho^{2}e^{\alpha E}]^{-3/2}
\right\},\nonumber\\
&&\hspace{-5mm}
H=\frac{\beta\varepsilon^{1/2}_{1}}{Z_{0}}\,e^{\alpha E}\rho\,{\rm Re}\left\{i[(1-i\theta)^{2}+\rho^{2}e^{\alpha E}]^{-3/2}
\right\}.\label{eq15}
\end{eqnarray}
Hereafter, $\theta=\tau+\alpha Z_{0}\rho H/(2\sqrt{\varepsilon_{1}})$. Once the
solution of nonlinear equations~(\ref{eq1}) is found, it is a simple matter to examine what
initial conditions are satisfied by it. Substituting $\tau=0$ into Eqs.~(\ref{eq15}),
we get
\begin{equation}
E=\beta[1+\rho^{2}\exp(\alpha E)]^{-3/2},\quad
H\equiv 0.\label{eq16}
\end{equation}
As a result, the implicit functions $E(\rho,\tau)$ and $H(\rho,\tau)$
determined by formulas~(\ref{eq15}) give the exact solution of
the Cauchy problem for system~(1) under initial conditions~(\ref{eq16}).
\begin{figure}[h]
\includegraphics{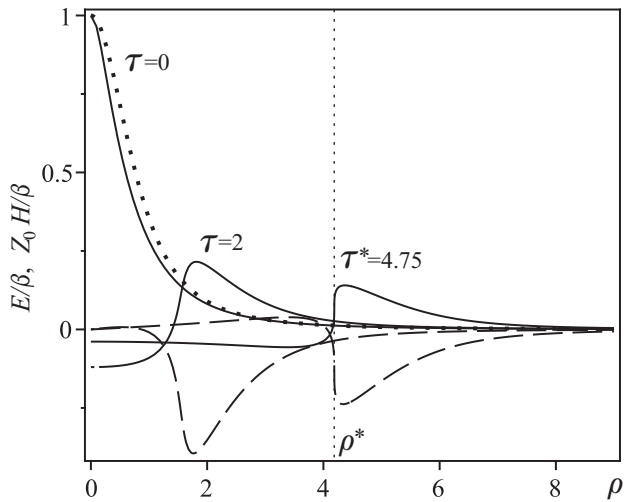}% Here is how to import EPS art
\caption{Radial distributions of the electric field (solid curves) and magnetic
field (dashed curves) at various times $\tau$. The dotted curve corresponds
to ${\cal E}(\rho,0)$ in Eqs.~(\ref{eq13}).}
\end{figure}

Fig.\,1 shows results of numerical calculations of $E$ and $H$ by
formulas~(\ref{eq15}) in the case where $\alpha\beta=1$ and
$\varepsilon_{1}=2$. The electric and magnetic fields as functions of
the coordinate $\rho$ are shown by the solid and dashed curves,
respectively, at various times $\tau$. The dotted curve corresponds
to the initial distribution~(\ref{eq13}) of the field $E$ in the auxiliary
linear problem. It is evident that even for a sufficiently
strong nonlinearity ($\alpha\beta=1$), the difference between the initial
conditions~(\ref{eq13}) and~(\ref{eq16}) is small. Fig.\,1 shows that the wave-profile part for which $E/\beta$ increases in the wave propagation direction becomes
steeper with time, thereby exhibiting the so-called self-steepening.
As a result, inflection of the wave profile occurs at a certain point
$\rho=\rho^{*}$ at the time instant $\tau=\tau^{*}$. For $\tau>\tau^{*}$,
three different values of both $E$ and $H$, which satisfy system~(\ref{eq15}),
correspond to one value of $\rho$, so that the curves of $E$ and $H$
become ambiguous for a given $\tau$. Due to this fact, discontinuities
of the wave components appear at the inflection point~\cite{Lan},
which corresponds to the formation of a cylindrical shock electromagnetic
wave. Upon appearance of discontinuities, the solution in form~(\ref{eq15})
ceases to be suitable.

The appearance of discontinuities of electromagnetic quantities results
from neglecting dispersion. Its influence leads to that the fields vary
continuously  under actual conditions. In this case, by a shock wave
one should understand a sufficiently rapid variation in the field
components on a certain moving interval. The thickness of this interval
(shock front) sets so as to enable the polarization of the medium
to switch from one value to the other.

{\em Boundary value problem.} Now consider a cavity resonator,
which is a perfectly conducting circular cylinder of radius $a$ and
height $L$. We assume that the $z$ axis is aligned with the
cavity axis and the perfectly conducting end walls of the cavity
are at $z=0$ and $z=L$. In the case where the cavity resonator is filled
with a linear medium having the permittivity
$\varepsilon=\epsilon_{0}\varepsilon_{1}={\rm const}$, $E_{0n0}$ (${\rm TM}_{0n0}$)
modes can exist in the cavity. The $E_z$ and $H_\phi$ components,
which are nonzero in these modes, are independent of $\phi$ and $z$, and are
described by the solutions of system~(\ref{eq10}) with the boundary
conditions
\begin{equation}
\left.E_{z}\right|_{r=a}\equiv{\cal E}(1,\tau)=0,\;\;
|\left.E_{z}\right|_{r=0}|\equiv |{\cal E}(0,\tau)|<\infty.\label{eq17}
\end{equation}
Such solutions are well-known and their derivation can be found elsewhere~\cite{Jac}.
Substituting the functions ${\cal E}$ and ${\cal H}$, which describe the
$E_{0n0}$ modes, into Eqs.\,(\ref{eq12}), we obtain the solution
of nonlinear equations~(\ref{eq1}) in the form
\begin{eqnarray}
&&\hspace{-5mm}
E=A\,J_{0}(\kappa_{n}\,\rho\,e^{\alpha E/2})\,\cos(\kappa_{n}\theta),\nonumber\\
&&\hspace{-5mm}
H=-AZ^{-1}_{0}\varepsilon^{1/2}_{1}\,e^{\alpha E/2}
J_{1}(\kappa_{n}\,\rho\,e^{\alpha E/2})\,\sin(\kappa_{n}\theta),\label{eq18}
\end{eqnarray}
where $J_{m}$ is a Bessel function of the first kind of order $m$, $\kappa_{n}$ is
the $n$th root of the equation $J_{0}(\kappa)=0$, and $A$ is an arbitrary
amplitude factor. Note that the field $E=0$ satisfies the transcendental
equations~(\ref{eq18}) for any $\tau$ if $\rho=r/a=1$. Therefore, the boundary
conditions~(\ref{eq17}) remain valid for the implicit function $E(\rho,\tau)$
defined by Eqs.~(\ref{eq18}). Thus, formulas~(\ref{eq18}) yield an exact solution of
the nonlinear boundary value problem for system~(\ref{eq1}) under conditions~(\ref{eq17})
and describe free electromagnetic oscillations in a cylindrical cavity
filled with a nonlinear medium.

Implicit solutions $E$ and $H$ given by Eqs.~(\ref{eq18}) and corresponding to a certain index
$n$ are periodic functions of time $t$ with period
$T_{n}=2\pi/\omega_{n}$, where
$\omega_{n}=\kappa_{n}(\epsilon_{0}\varepsilon_{1}\mu_{0})^{-1/2}a^{-1}$ is
an eigenfrequency of the $E_{0n0}$ mode. Along with the fundamental
frequency $\omega_{n}$ for each $n$, the Fourier time series expansions
of the functions $E$ and $H$ also contain terms at the
multiple frequencies $l\omega_{n}$, where $l$ is an integer. The contribution
of harmonics with $l\geq 2$ determines the role of nonlinear effects
which manifest themselves as deviations of the quantities $E$ and $H$
from their values corresponding to the
$E_{0n0}$ mode in a cavity with
$\varepsilon=\epsilon_{0}\varepsilon_{1}={\rm const}$
in the linear case ($\alpha=0$).
\begin{figure}[h]
\includegraphics{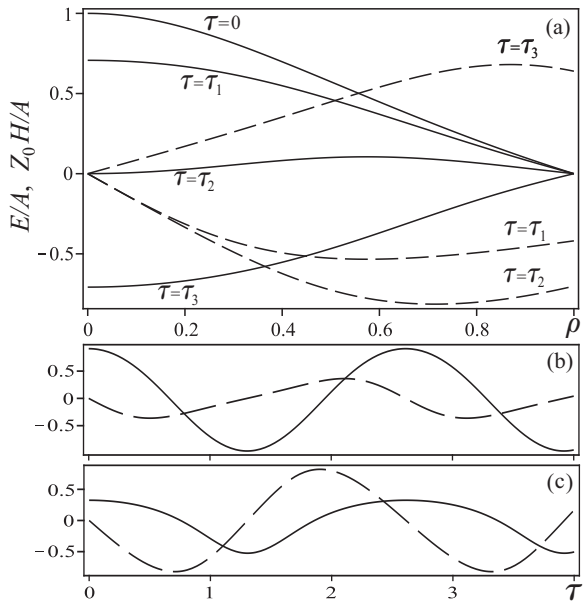}% Here is how to import EPS art
\caption{(a)
Electric and magnetic fields as functions of $\rho$ (solid and dashed
lines, respectively) in
the $n=1$ mode at times $\tau_{1}=\pi/(4\kappa_{1})$, $\tau_{2}=\pi/(2\kappa_{1})$,
and $\tau_{3}=5\pi/(4\kappa_{1})$.
Oscillograms of the fields at (b) $\rho=0.2$ and (c) $\rho=0.7$.}
\end{figure}
\begin{figure}
\includegraphics{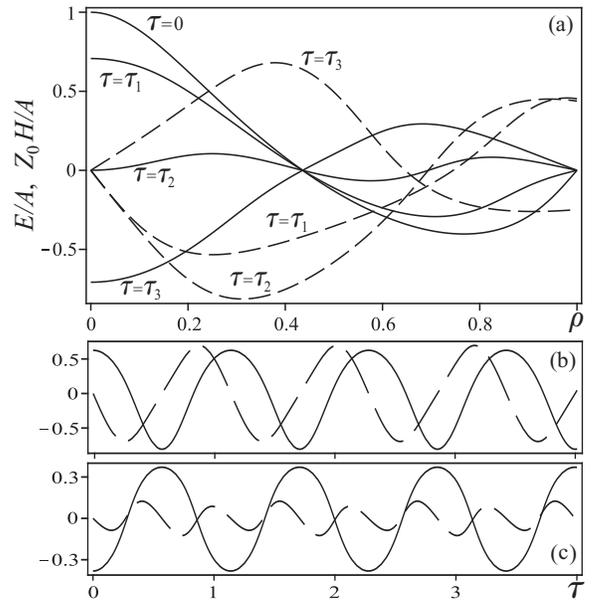}% Here is how to import EPS art
\caption{(a) The same as in Fig.\,2(a), but for
the $n=2$ mode at $\tau_{1}=\pi/(4\kappa_{2})$, $\tau_{2}=\pi/(2\kappa_{2})$,
and $\tau_{3}=5\pi/(4\kappa_{2})$. Field oscillograms at (b) $\rho=0.2$ and (c) $\rho=0.7$.}
\end{figure}
Let us now turn to results of calculations of the quantities $E$ and $H$
by formulas~(\ref{eq18}). Fig.\,2(a) shows snapshots of the normalized field
components $E/A$ and $Z_{0}H/A$ in the lowest mode
($n=1$ and $\kappa_{1}\simeq 2.4$) as functions of $\rho$ at fixed instants
of time $\tau$. Figs.\,2(b) and 2(c) show the oscillograms of the
field components at two points $\rho=0.2$ and $\rho=0.7$ for
$n=1$. Similar curves for a mode with $n=2$ and $\kappa_{2}\simeq 5.5$
are presented in Fig.\,3. Figs.\,2 and 3 were plotted for $\alpha A=0.5$ and
$\varepsilon_{1}=2$.
The presented plots show that the nonlinear
effects become more pronounced with increasing $n$ and depend significantly
on the coordinate $\rho$, i.e., location of the observation point inside
the cavity. For example, the field oscillograms in Fig.\,3(b) are
analogous to those in the linear case. However, in Fig.\,3(c) we see that
the field $E$
varies at the frequency $\omega_{2}$, while the field $H$, at the
second harmonic $2\omega_{2}$. For the higher modes with $n>n^{*}$, where
$n^{*}$ is an integer depending on the parameter $\alpha A$, the functions
$E(\rho,\tau)$ and $H(\rho,\tau)$ determined by~(\ref{eq18}) become ambiguous in
a certain domain of values of the variables $\rho$ and $\tau$. Since
such behavior is not physically admissible, one should expect field
discontinuities at the ambiguity points. The time dependences $E(\tau)$
and $H(\tau)$ can then be discontinuous (relaxation) oscillations,
and the solutions~(\ref{eq18}) obtained without allowance for dispersion become
inapplicable. However, it is important to emphasize that for weak
nonlinearity ($|\alpha A|\ll 1$), the number $n^{*}$ is large (e.g.,
$n^{*}=9$ for $\alpha A=0.5$) and solutions~(\ref{eq18}) for $n<n^{*}$ are
single-valued continuous functions of coordinates and time. Due to this
fact, the exact solutions found seem to be of great practical
interest and can be used for analysis of, e.g., ferroelectric
resonators.

In conclusion, we note that the proposed method makes it to possible to easily generate various physically interesting solutions of nonlinear system~(\ref{eq1}), starting from the corresponding solutions of linear field equations. Therefore, this method has significant advantages over the direct numerical solution of that system.

This work was supported by the RFBR (project no. 09--02--00164-a)
and the Russian Federal Program ``Kadry.''

\bibliography{Petrov_Kudrin_bibl}

\end{document}